\documentclass[conference]{IEEEtran}
\IEEEoverridecommandlockouts

\usepackage{cite}
\usepackage{amsmath,amssymb,amsfonts,bm}
\usepackage{todonotes}

\usepackage{hyperref}
\hypersetup{
    colorlinks=true,
    linkcolor=blue,
    filecolor=magenta,      
    urlcolor=black,
}
\usepackage{algorithmic}
\usepackage[boxed]{algorithm2e}
\usepackage{graphicx}
\usepackage{subfig}
\usepackage{textcomp}
\usepackage{xcolor}
\def\BibTeX{{\rm B\kern-.05em{\sc i\kern-.025em b}\kern-.08em
    T\kern-.1667em\lower.7ex\hbox{E}\kern-.125emX}}

\newcommand{\x}{\mathbf{x}}
\newcommand{\s}{\mathbf{s}}
\newcommand{\y}{\mathbf{y}}

\newcommand{\Ph}{\boldsymbol{\Phi}}
\newcommand{\ph}{\boldsymbol{\phi}}

\newcommand{\R}{\mathbb{R}}
\newcommand{\rr}{\boldsymbol{r}}

\begin{document}

\title{DeepMP for  Non\textendash Negative  Sparse Decomposition}

\author{\quad \quad \quad\quad Konstantinos A. Voulgaris \quad Mike E. Davies \quad Mehrdad Yaghoobi 
\\ \centerline{ Institute for Digital Communications, the University of Edinburgh, EH9 3JL, UK} \\ $\left \{\textrm{Konstantinos.Voulgaris, Mike.Davies, m.yaghoobi-vaighan}\right \}$@ed.ac.uk 
}

\maketitle

\begin{abstract}
Non\textendash negative signals form an important class of sparse signals. Many algorithms have already been
proposed to recover such non-negative representations, where
greedy and convex relaxed algorithms are among the most
popular methods. The greedy techniques are low computational cost algorithms, which have also been modified
to incorporate the non-negativity of the representations. One
such modification has been proposed for  Matching
Pursuit (MP)  based algorithms, which first chooses positive coefficients and uses
a non-negative optimisation technique that guarantees the non\textendash negativity of the coefficients. The performance of greedy algorithms, like all non\textendash exhaustive search methods, suffer from high coherence with the linear generative model, called the dictionary. We here first reformulate the non\textendash negative matching pursuit algorithm in the form of a deep neural network.  We then show that the proposed model after training yields a significant improvement in terms of exact recovery performance, compared to other non\textendash trained greedy algorithms, while keeping the complexity low.  
\end{abstract}


\section{Introduction}
Sparse coding is the problem of reconstructing input vectors using a linear combination of an overcomplete  family basis vectors  with sparse coefficients. It has become extremely popular for extracting features from raw data,  particularly when the dictionary of basis vectors is learned from unlabeled data. There exist several unsupervised learning methods that have been proposed to learn the dictionary. Applications of sparse coding may be found in fields such as visual  neuroscience\cite{sc1},\cite{sc2} and image restoration \cite{sc3},\cite{sc4}. A major problem with these methodologies is that the inference algorithm is somewhat expensive, prohibiting real\textendash time applications. 

Let the signal of interest be $\mathbf{y} \in \mathbb{R}^{M}$ and a dictionary of elements $\mathbf{\Phi} \in \mathbb{R}^{M\times N}$ be given. The linear sparse approximation can be formulated as finding the sparsest $\mathbf{x} \in \mathbb{R}^{N}$, $M<N$, i.e having the minimum number of non\textendash zero elements, as follows:
\begin{equation}
\mathbf{y}\approx \mathbf{\Phi}\mathbf{x}
\end{equation}
The greedy sparse approximation algorithms are in general known for  low computational costs, suitable for real\textendash time and large scale sparse approximations. The   Matching Pursuit (MP)  \cite{c1}, algorithm is  introduced, which approximately solve the following problem: 
\begin{equation}\label{eq:ref}
\tilde{\mathbf{x}}:= \textrm{argmin}_{x_{s}} ||\mathbf{y}-\mathbf{\Phi}_s\mathbf{x}_s||_2,\newline
\end{equation}
where $s$ is a subset of all atoms with cardinality $k$.

There are many applications for which the coefficient vectors are  not only sparse, but also non\textendash negative. Spectral and multi\textendash spectral unmixing, \cite{c4},\cite{c5},  microarray analysis \cite{c6} and  Raman spectral deconvolution \cite{c7} are a few examples.

The original implementation  of MP has been modified in order to adopt the algorithm to the non\textendash negativity setting. Essentially the original minimization problem introduced in \eqref{eq:ref} is reformulated by adding a constraint that guarantees the non\textendash negativity of the coefficients and takes the following form:
\begin{equation}
\begin{aligned}
& \tilde{\mathbf{x}}:= & & \textrm{argmin}_{x\geq 0} ||\mathbf{y}-\mathbf{\Phi} \mathbf{x}||_2 \\& \text{} & & ||\textbf{x}||_0\leq j
\end{aligned}
\end{equation}\label{eq:ref_1}
 where $||\cdot||_0$ measures the number of non\textendash zero elements.

MP incrementally builds up $\x$ with respect to the set of columns of $\mathbf{\Phi}$, $s_a$. A known fact about MP algorithms is  that the obtained solution can be an approximation of the input signal $\mathbf{y}$. The acquired support set $s_a = \left\{i: \phi_i \in \Phi\right\}$ is then different  to the ground truth support set $s_g$. Hence, given a $k$\textendash sparse signal $\mathbf{x}$, it is quite frequent to have $|s_a\cap s_g|< k$, particularly  when the atoms in the dictionary are correlated, i.e coherent dictionary.

The authors in \cite{c3} introduced  the Learned Iterative and Thresholding Algorithm (LISTA). Under the assumption that the basis vectors have been trained and are being fixed, the core idea is to train a parameterized encoder function to predict the optimal sparse code. A key advantage of this physical\textendash model based framework  is that it has a predetermined\textendash complexity and can be used to approximate sparse codes with a fixed computational cost and a prescribed expected error that makes it appealing for real\textendash time applications which is the main focus of this study. Recent advances in the LISTA framework \cite{LISTA_con} introduced the theoretical conditions upon the convergence of the algorithm, while in \cite{LISTA_ana} the authors demonstrated that by  following an analytical approach, rather than a learned approach, the network retains its optimal linear convergence. They later introduce an acceleration technique in the training procedure, given that the number of training parameters is significantly reduced. 

Inspired by  the unfolding idea introduced by the LISTA framework to reformulate the convex optimization algorithms with deep neural networks (DNN), we here introduce a variation of the original NNMP, as introduced in \textbf{Algorithm 1}, called DeepMP \footnote[1]{code available in:  \url{https://github.com/dinosvoul/Deep-Matching-Pursuit}} as a data adaptive and bounded complexity algorithm. Our preliminary goal is  to introduce a novel framework for  non\textendash negative sparse approximation which outperforms the existing ones in terms of accuracy, maintaining the computational cost. Nevertheless the canonical linear approach followed at the selection step of NNMP even though it is simple computationally  is not data adaptive and therefore not flexible. The DeepMP approach has more flexible nature  introducing a higher degree of freedom at the selection step representation underlying complex nature of data  i.e human handwriting. 

\section{Deep Matching Pursuit}

\begin{algorithm}[t!]
	\small
	\caption{Non-Negative Matching Pursuit algorithm (NNMP)}\label{alg:nnmp}
	\begin{algorithmic}[1]
		\STATE \textbf{initialisation:} $s = \emptyset$, $k = 0$ and $\mathbf{r}_0 = \y$ 
		\WHILE {$k < K \ \& \ \max(\Ph^T \mathbf{r}_k) > 0 $} \label{alg:cnnompmax}
		\STATE $\s_k = \mathbf{0}$
		\STATE \color{red}{$(\zeta,\iota) \leftarrow \max(\Ph^T \mathbf{r}_k)$} \color{black}
		\STATE $\s_k[\iota] = \zeta$
		\STATE \color{red}{$\mathbf{r}_{k+1} \leftarrow \mathcal{P} \{\mathbf{r}_{k} - \zeta\ph_\iota \}$} \color{black} \label{alg:nnmpr}
		\STATE $k \leftarrow k+1$
		\ENDWHILE
		\STATE $\x \leftarrow \sum_{k} \s_k$
	\end{algorithmic}
\end{algorithm}

\begin{figure}[t!]
	\centering
	\includegraphics[width=8.8 cm]{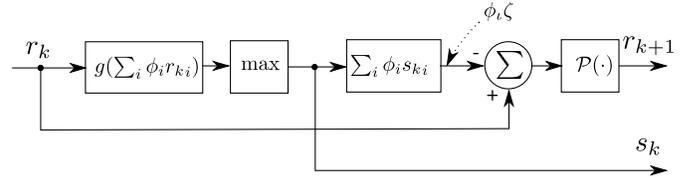}
	\caption{\small One step of non-negative matching pursuit algorithm. $r_k$ and $s_k$ are respectively the residual and the selected index at the $k^{th}$ step of algorithm.}\label{fig:mp1s}
	\vspace{-3mm}
\end{figure}

In this section we introduce the modification in the standard NNMP structure which is introduced in Algorithm 1.  Starting with the measurement $\y$ as the current residual signal $\rr_{\small k}|_{\tiny k = 0}$, the main steps are: a) finding the best matched atom $\ph_{\small k}$ to $\rr_{\small k}$, and 2) updating the residual $\rr_{\small k}$ by subtracting the contribution of selected atom. The operator $\mathcal{P}$ can be identity or the projection onto the positive orthant which is done such that the framework  is consistent with the non\textendash negative setting. A flowchart diagram of each NNMP iteration has been shown in Figure \ref{fig:mp1s}, where "max" operator simply keep the largest component of the input and zero out the rest, and $\s_k$ is the 1-sparse vector with the appropriate coefficient on its support. The "max" operator is called "hard-max" operation here, which is the projection onto the best one-sparse set, also known as the 1-sparse hard-thresholding \cite{c2}.

\begin{figure}[t!]
	\centering
	\includegraphics[width=8.8 cm]{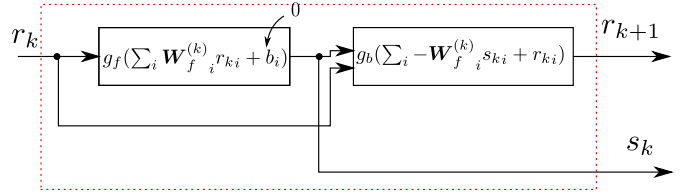}
	\caption{\small Representation of one step of NNMP algorithm in the form of a two layer neural network with a skip connection. $g_f$ and $g_b$ are respectively hard-max and linear/ReLU activation functions.}\label{fig:deepmp1s}
	\vspace{-3mm}
\end{figure}

 The NNMP steps are reformulated and the dictionary is replaced with the weight matrices of the same size $\boldsymbol{W}_{f}^{\tiny (k)}$ and $\boldsymbol{W}_{b}^{\tiny (k)}$ at iteration $k$. We then get the following (non-linear) system of equations,
\begin{equation}
\left( \begin{split}
\rr_{\tiny k+1} \\ 
\x_{\tiny k+1}
\end{split}
\right) 
= g_{\tiny b}\left(
\begin{split}
-\boldsymbol{W}_{b}^{\tiny (k)} g_{\tiny f}({\boldsymbol{W}_{f}^{\tiny (k)}}^T \rr_{\tiny k}) + \rr_{\tiny k} \\
g_{\tiny f}({\boldsymbol{W}_{f}^{\tiny (k)}}^T \rr_{\tiny k})  + \x_{\tiny k}
\end{split}
\right),
\end{equation}
where $g_f$ and $g_b$ are respectively forward and backward functions which are hard-max and $\mathcal{P}$. Such a model can be represented as two layers of a neural network model per single iteration of the algorithm. By concatenation of $K$ blocks of Figure \ref{fig:deepmp1s}. The depth of the network then varies depending on the sparsity of the signal $k$, and can be represented as the concatenation of $k$ blocks of Figure 3. The network takes then the form of a DNN  of $2 K$ layers.  $\x$ can then be reconstructed by superposition of $\s_k$s.

This structure provides a framework for sparse approximations, if we train the weight matrices using the backpropagation algorithm. Such a variational learning method with non-differentiable activations is possible using a surrogate function optimization method, \textit{e.g.} \cite{c9} for ReLU. As one candidate solution is $\boldsymbol{W}_{f}^{\tiny (k)} = \boldsymbol{W}_{b}^{\tiny (k)} = \Ph, \forall k$, the network will at least work as good as NNMP, if the learning is successful. Within our work we modify step 4 of NNMP by replacing the original library $\mathbf{\Phi}$ with $\boldsymbol{W}_{b}^{\tiny (k)}$ while in step 6 $\boldsymbol{W}_{f}^{\tiny (k)}=\mathbf{\Phi}$. At the particular step we introduce the non\textendash linear activation function $\mathcal{P}$ with the aim to project the residual vector $\rr$ on the positive orthant.

Essentially the key change in the NNMP structure is the more flexible approach  in step 4 of the algorithm. A successful decomposition of the input signal $\mathbf{y}$ depends on the proper selection of the candidate atom. A wrong selection can have a direct impact on the minimization problem introduced in \eqref{eq:ref}. By training DeepMP we  generate different copies of $\boldsymbol{W}_{b}^{\tiny (k)}$  over the $k$ layers of the network. We aim to have an approach that prevents  misclassifications and potentially  results in zero error for a fixed number of iterations and no noise in the input. Our expectation relies on the fact that DeepMP utilizes a higher degree of freedom in the selection step of the algorithm. In particular, the canonical approach of NNMP represents a model  that consists of a fixed number of $MN$ parameters. This kind of approach may be simple, but not flexible. On the other hand the DeepMP model consists of a number of parameters which scales linearly over the  network layers and  results in the total to a number of $KMN$ parameters at the selection step. In practice this means that the higher the sparsity of the signal $K$ the higher degree of flexibility introduced in the framework, i.e the capacity of the DNN, which improves the performance compared to canonical NNMP overall.

From classification point of view the DeepMP framework actually performs a multilabel classification task by decomposing the input signal $\mathbf{y}$ with respect to the corresponding classes, and using the categorical cross\textendash entropy loss function,
\begin{equation}\label{loss}
    H(p,q)= -\sum_{j=1}^{|\Phi|} \mathbf{1}_{s_a}(j)\textrm{log}\ p(j,i) 
\end{equation}
where $i$ corresponds to the $i$\textendash th sample, $j$ to the index of the atom and $ \mathbf{1}_{s_a}:I \rightarrow \left\{0,1\right\}$ is the indicator function, defined as:

\[ 
\mathbf{1}_{s_a}(j)= \left\{
\begin{array}{ll}
      1\ \textrm{if} \ j \in s_a   \\
     0 \ \textrm{if} \ j \not\in s_a 
\end{array} 
\right. 
\]

\subsection{Sparse Signal Decomposition}
The main motivation for introducing deep learning approach  is to introduce  more flexibility approach in the selection rule of the MP type algorithms. We aim to improve the prediction rate on the support set and eventually reduce the residual error compared to the standard MP framework. 

Considering the set of sparse signals which are the main focus of the current work. The main goal of the decomposition algorithm is to identify the atoms which build up the input signal $\mathbf{y}$ with non\textendash negative  weights as follows: 
\begin{equation}\label{eq:mix}
    \mathbf{y}=\sum_{l=1}^{k} a_w \boldsymbol\phi_i .
\end{equation}
 with $a_w \sim U[0,1]$, where $ U[0,1]$ stands for the uniform distribution with 0 mean and unit variance. 

The overall process can then be represented as an iterative algorithm. A common phenomenon that frequently takes place during the decomposition is the selection of unrelated atoms in the support set $s_a$, over the iterations of the algorithm.

A  reason for this phenomenon occurs, is related to the similarity between the atoms  $\pmb{\phi}_i$. In cases where the algorithm  operates over a point cloud where the constituent atoms are highly coherent with each other, the algorithm may select a neighboring atom instead  of  the ground truth atoms in $s_g$, i.e $i \in s_a$ while $i \not\in s_g$. Coherence measures the  maximum similarity  between two distinctive atoms of $\mathbf{\Phi}$. Given a pair of points $\pmb\phi_i, \pmb\phi_j \in \boldsymbol\Phi$ where $i\neq j$, the coherence can be formulated as follows:

\begin{equation}
    \mu(\boldsymbol\Phi)=\max_{i\neq j} \frac{|\langle \pmb\phi_{i},\pmb\phi_j \rangle|}{||\pmb{\phi}_i||_2\cdot||\pmb{\phi}_j||_2}
\end{equation}
 
 where  $||\cdot||_2$ indicates the Euclidean norm. By introducing the matrix $\boldsymbol{W}_{b}^{\tiny (k)}$ at the selection step of the algorithm, we are aiming for the points to be represented in a way that the mutual coherence of the points will decrease. In that sense by training the network we are expecting that the coherence of the corresponding representation  $\boldsymbol{W}_{b}^{\tiny (k)}$ yields an outcome where ideally $\mu(\boldsymbol\Phi)\geq \mu(\mathbf{W}_b^{(k)})$, where
 $\mu(\boldsymbol\Phi)$ and  $\mu(\mathbf{W}_b^{(k)})$ are respectively the coherence in  $\boldsymbol\Phi$ and $\mathbf{W}_b^{(k)}$.

\section{Experiments}
Within the current section we evaluate the performance of DeepMP by some simulations. In order to evaluate the performance of DeepMP 
we are considering two datasets;  a synthetic dataset 
$\boldsymbol\Phi \in \R_{+}^{d \times N}$. The dictionary was randomly generated with an \textit{i.i.d.} normal distribution and  then projected onto the positive orthant and column normalised. A real dataset of \textit{Raman} spectra, where each of the spectras consists of 503 wavenumbers that lay within the range of  $306$ to $1249\ \textrm{cm}^{-1}$, provided by \cite{data} . 
We perform a number of 150000 trials for each dataset   where only $\boldsymbol{W}_{f}^{\tiny (k)}$s were trained in the $M$-space while $\boldsymbol{W}_{b}^{\tiny (k)}=\mathbf{\Phi}$. This essentially means that we only train the weights that correspond to the selection step of the algorithm while the weights that correspond to the update step are kept fixed.  This is done because we address DeepMP as a solver for the standard  non\textendash negative least squares problem as introduced in (3). In case where the weights in the update step are also trained the cost function of the problem is reformulated as follows: $ \textrm{argmin}_{\mathbf{x}_{s}\geq 0} ||\mathbf{y}-\mathbf{W}_{b,s}{\tiny (k)} \mathbf{x}_s||_2$.  

The DeepMP framework is optimized using the   \textit{AdaBound}  algorithm \cite{ADABOUND}. More details about the datasets and the settings for the \textit{AdaBound} algorithm can be found in TABLE I.

\begin{table}[t!]
\centering
\begin{center}
\scalebox{1.17}{\begin{tabular}{|c|c|c|c|c|c|}
\hline
& N & M  &  $\textrm{lr}$ & $\textrm{final}\_\textrm{lr}$& epochs \\ \hline
Synthetic data & 30 & 200  & $1e-3$ & $0.1$&20  \\ \hline
\textit{Raman} library & 503 & 2521  & $1e-3$ & $0.1$ & 30 \\ \hline
\end{tabular}}
\end{center}\vspace{-0.25cm} \caption{}\label{tab:label}
\end{table}

As an evaluation metric for the exact recovery of the support set we are using the normalized \textit{Hamming distance} complement \cite{metric}. The metric is defined as in \eqref{eq:num}.
\begin{equation}\label{eq:num}
\begin{aligned}
H^\complement(s_a,s_g)= \sum_{n=1}^{N}1-\frac{ |s_a(n)-s_g(n)|}{k}
\end{aligned}
\end{equation}
 where $S_a$ is the support set acquired by the corresponding algorithm and $S_g$ the ground truth, $k$ is the sparsity level and $|\cdot|$  the cardinality operator. The performance on the reconstruction error for each sparsity $k$ is evaluated with respect to $\epsilon$ as follows:
\begin{equation}\label{eq:error}
  \epsilon(k)= \frac{1}{Z}\sum_{z=1}^{Z}\frac{||\mathbf{y}[z]-\mathbf{\Phi} \mathbf{x}[z]||_2}{||\mathbf{y}[z]||_2},
\end{equation}
 where $Z$ the number of realizations.

A basic expectation while training the selection step of the algorithm is the variation of the coherence in between $\mathbf{w}_i^{k},\mathbf{w}_j^{k}$ columns of $\mathbf{W}_{b}^{\tiny (k)}$. For the particular aspect of the problem we use the empirical cumulative distributed function (ECDF) as introduced in equation \eqref{eq:ecdf}:

\begin{equation}\label{eq:ecdf}
   \mu_{\textrm{ECDF}}(t)=  \frac{1}{\binom{|\Phi|}{2}} \sum_{i=1}^{M}\sum_{j\neq  i}^{M} \mu(\pmb\phi_i,\pmb\phi_j)\leq t.
\end{equation}

where $t \in [0,1]$.

\subsection{Results}

\begin{figure}[t!]%
    \centering
    \subfloat[Support Recovery]{{\includegraphics[width=3.9cm]{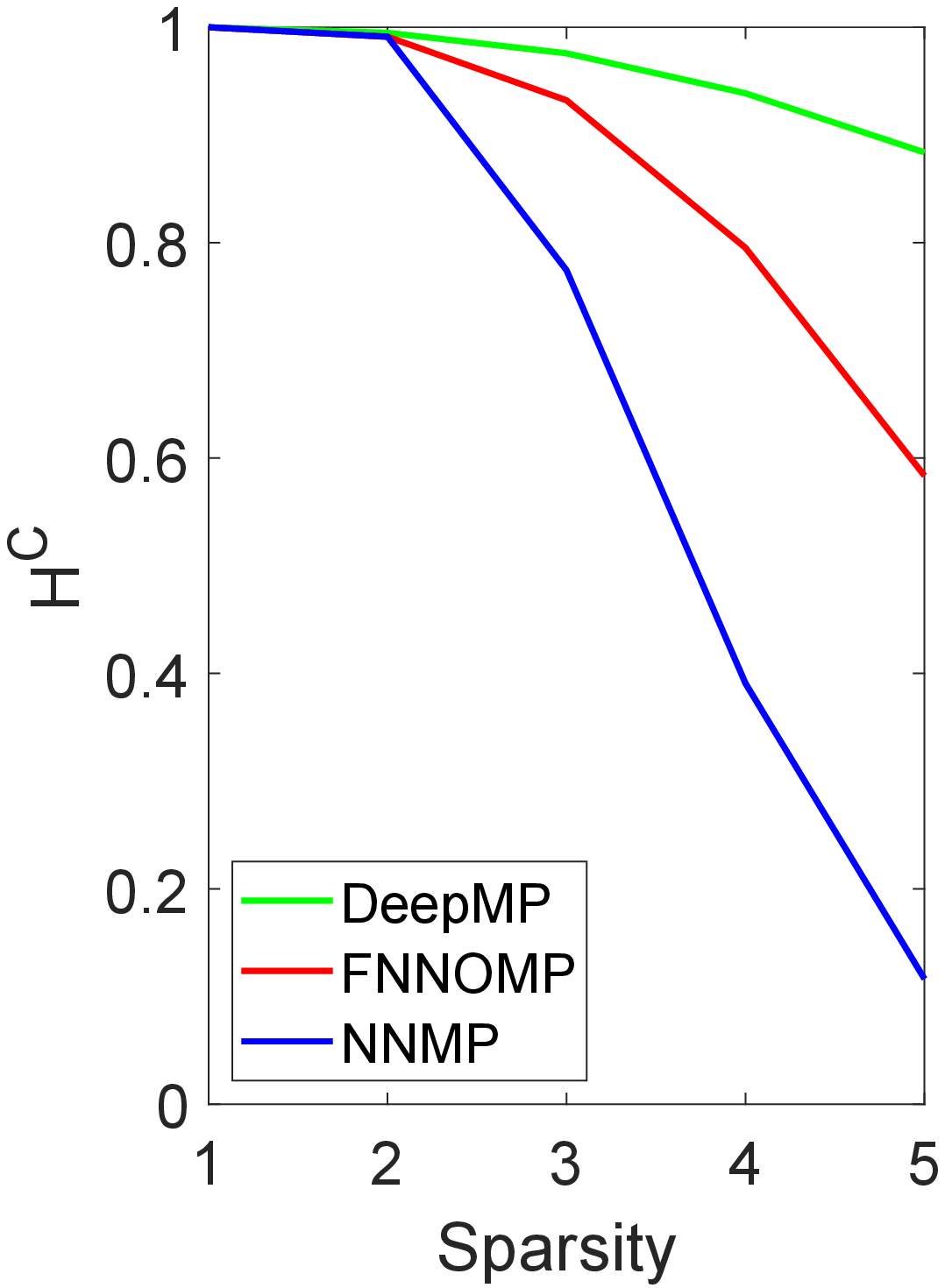} }}%
    \qquad
    \subfloat[Reconstruction error]{{\includegraphics[width=3.9cm]{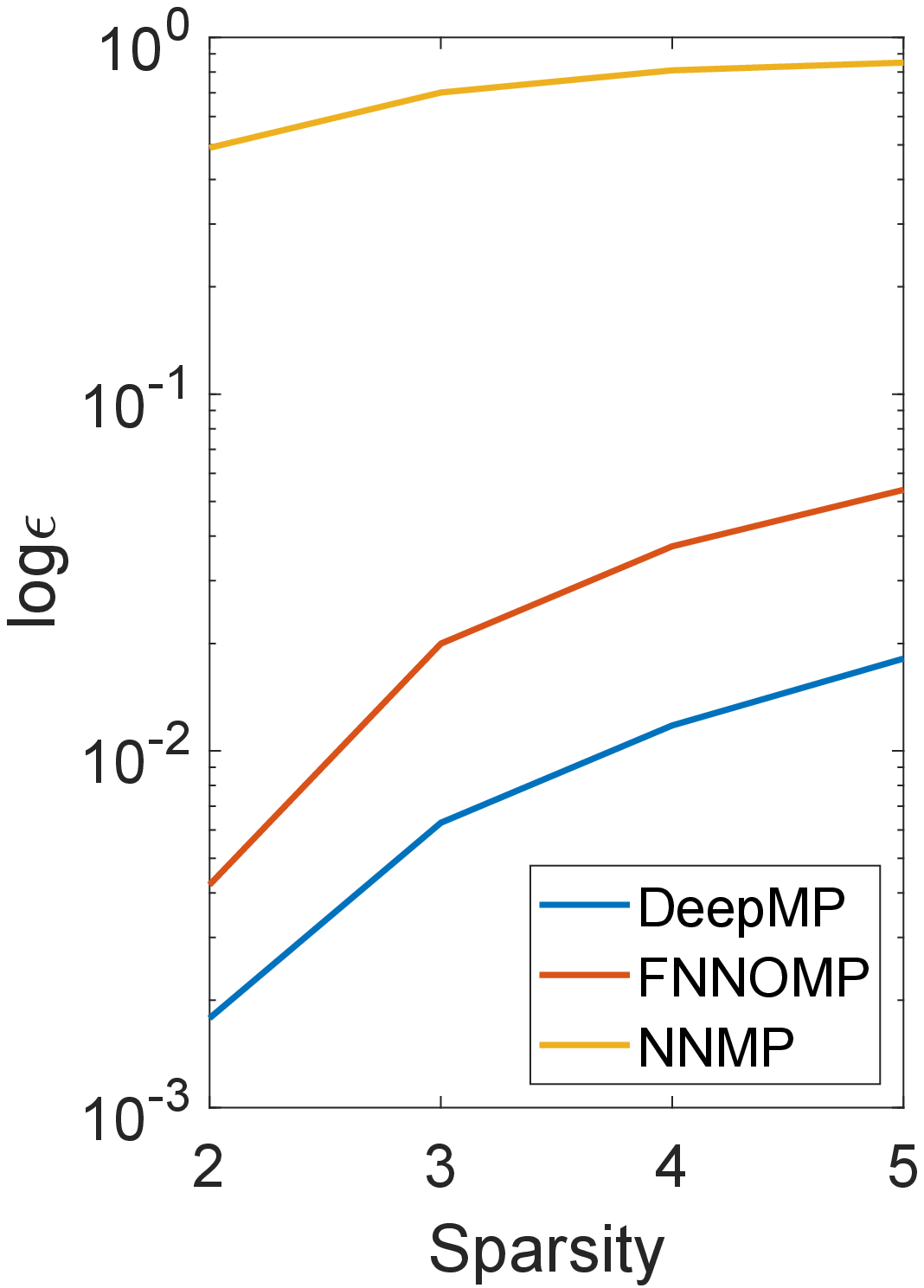} }}%
    \caption{The performance of  different MP frameworks with the \textit{Raman} dictionary.}%
    \label{fig:raman}%
\end{figure}

\begin{figure}[b!]%
    \centering
    \subfloat[Support Recovery]{{\includegraphics[width=3.9cm]{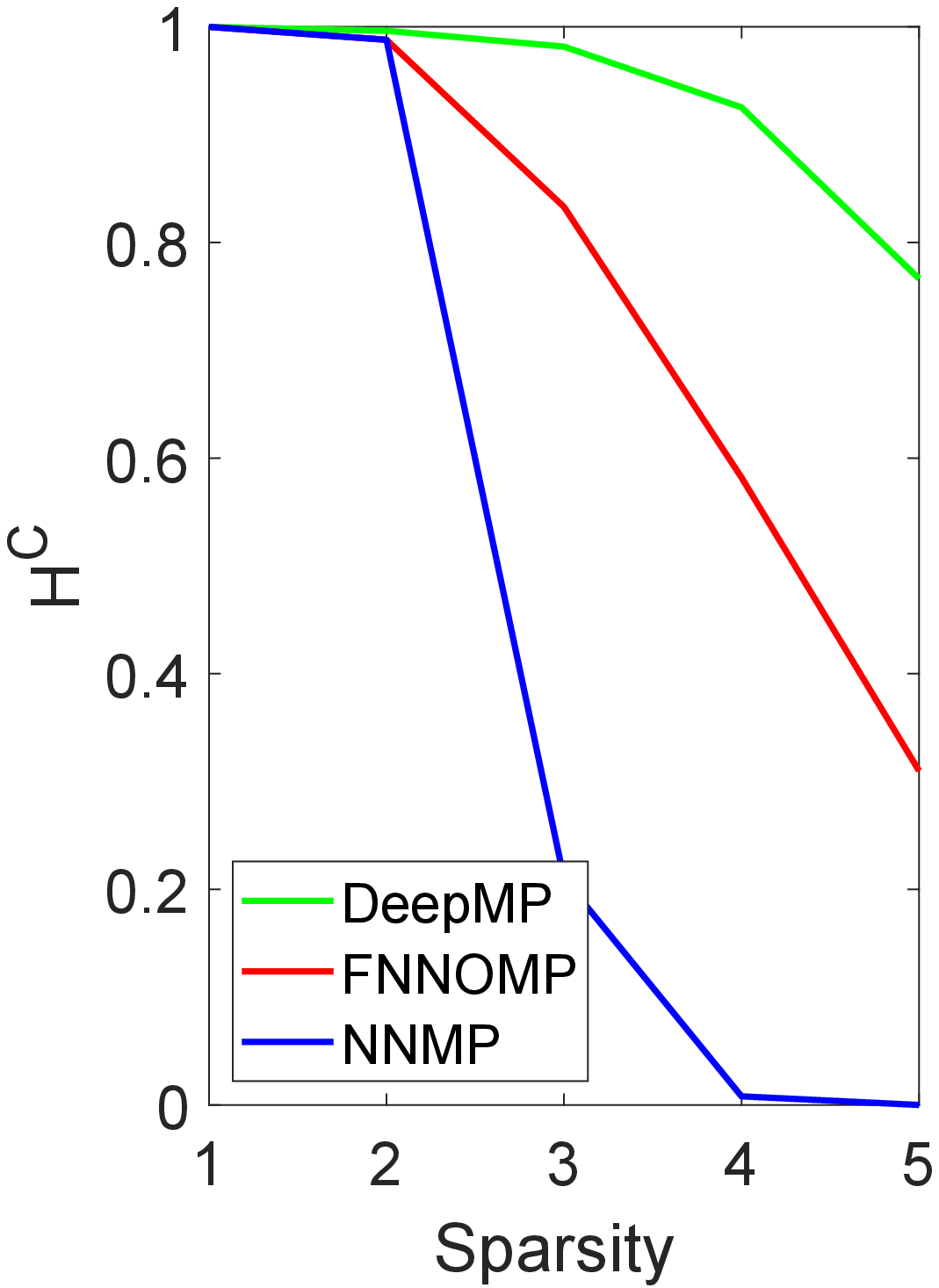} }}%
    \qquad
    \subfloat[Reconstruction error]{{\includegraphics[width=3.9cm]{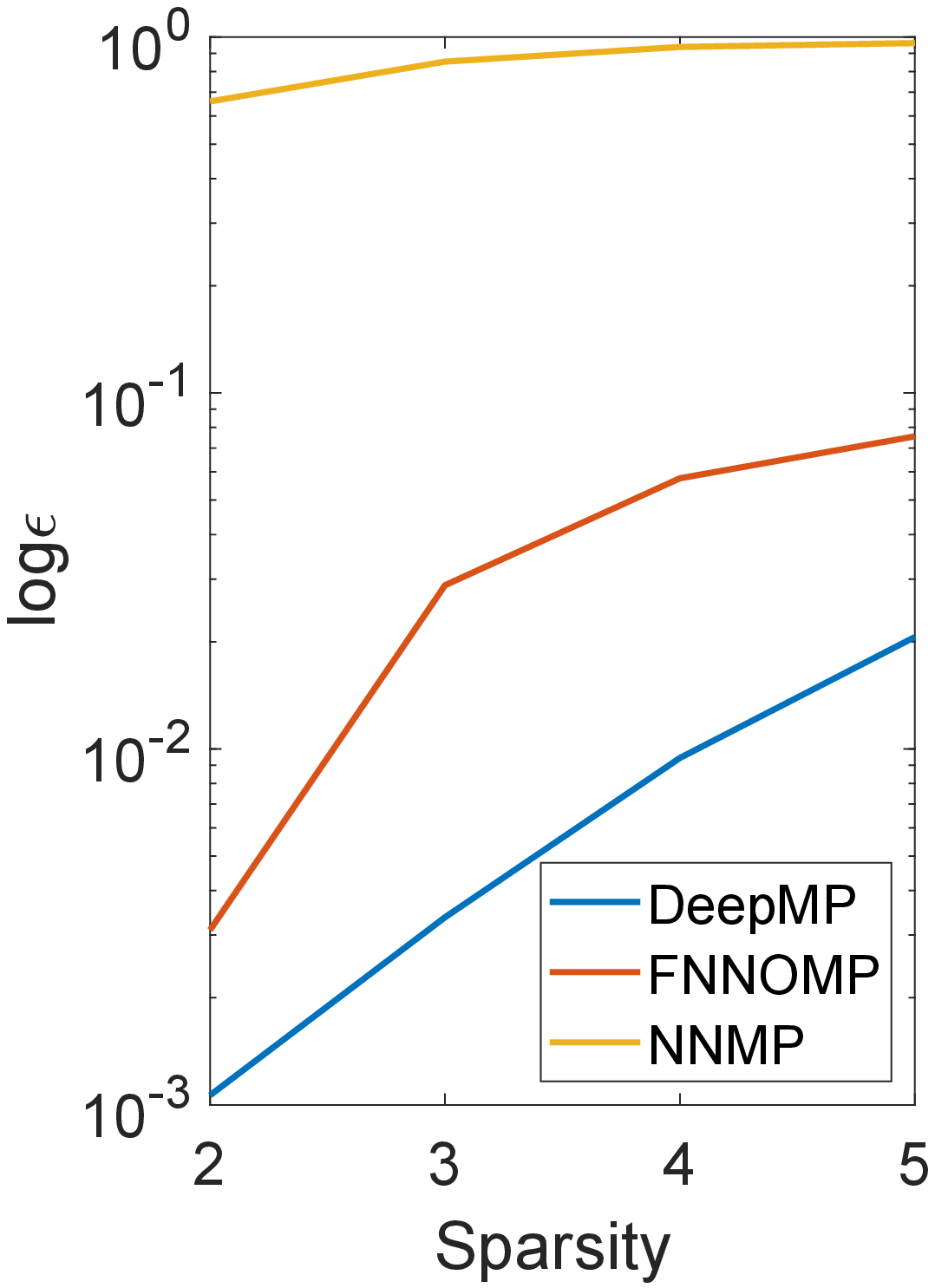} }}%
    \caption{The figure demonstrates the performance of the different MP frameworks for the Synthetic dictionary.}%
    \label{fig:synthetic}%
\end{figure}
 We here perform a simulation based  evaluation of the different MP frameworks. We are particularly interested in signals which are very sparse. Hence we  consider mixtures of signals $\mathbf{y}$ that consist of up to 5 atoms. From the perspective of the DeepMP framework this corresponds to concatenation of up to 5 different versions of the model with a varying depth over sparsity. These versions  are independent, i.e the 1st layer  is different from the one model to the other. The obtained results for the \textit{Raman} data and the synthetic data are demonstrated in figures \ref{fig:raman}
and \ref{fig:synthetic} accordingly. 

As it can be seen from the results, DeepMP outperforms the NNMP and FNNOMP\cite{c8} with respect to the Hamming distance complement, while FNNOMP significantly outperforms NNMP. This essentially means that the extra degrees of freedom  on the  selection step of   DeepMP introduces a more flexible approach which overall, leads to a better  exact recovery performance. The advantage of DeepMP  becomes  more significant over sparsity having less sparse signals, i.e larger $k$. Hence, despite the fact that the performance of all the MP frameworks decays with $K$, the flexibility of DeepMP leads to a slower decay over sparsity and hence getting a better performance compared to FNNOMP and NNMP. The $\epsilon$\textendash error results  also indicate that the improved exact recovery, leads to a better performance on the reconstruction of the input signal $\mathbf{y}$.

\begin{figure}[t!]%
    \centering
    \subfloat[\textit{Raman} dictionary]{{\includegraphics[width=8cm]{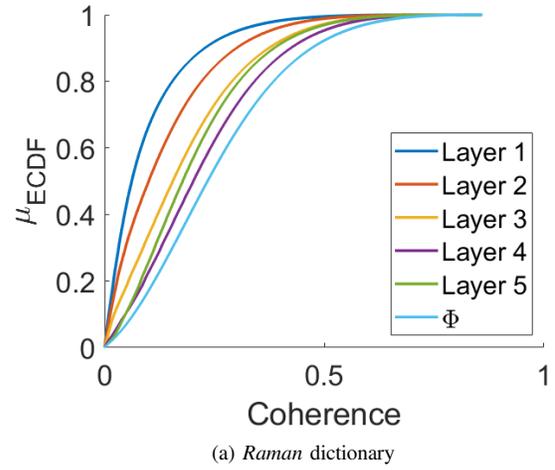} }}%
    \qquad
    \subfloat[Synthetic dictionary]{{\includegraphics[width=8cm]{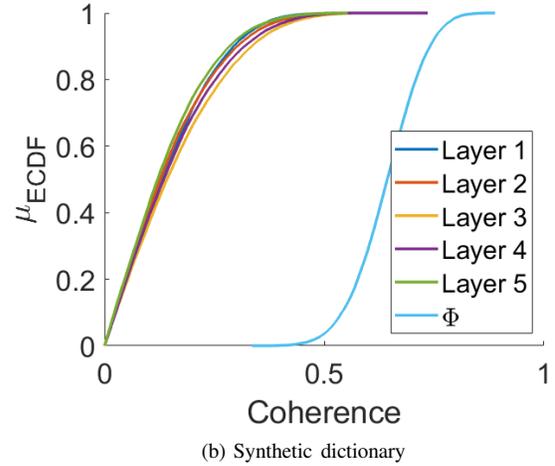} }}%
    \caption{The coherence of the Raman dictionary and the synthetic dictionary in comparison with the coherence of the weight matrices of DeepMP for $k=5$.}%
    \label{fig:coherence}%
\end{figure}

Despite demonstrated  good results using DeepMP, a  question  is why it outperforms NNMP and FNNOMP. While a rigorous answer to this question is left for the future, we demonstrate the $\mu_{\textrm{ECDF}}$'s of the two dictionaries and the trained  model trained for $k=5$  in figure \ref{fig:coherence}. As it can be seen from the results, the network  generates weight matrices with  reduced coherences compared to the original ones. In that sense, the corresponding point clouds consist of a set of atoms which are further apart the one to the other. This essentially means that DeepMP alters the underlying geometry of the selection step to avoid a misclassification. Given that the points are further apart, i.e smaller coherence in average,  the algorithm can easier pick the right atom without confusing it with its  neighbors.

\section{Conclusion\textendash Future work}

We here introduces DeepMP which is a novel framework for  non\textendash negative sparse decomposition. The main goal of the current work is to maintain the computational advantages of the standard MP algorithm while improving the performance of  signal approximation. The obtained results indicate that DeepMP outperforms the standard  MP approaches in terms of signal reconstruction. A future direction of the current work can be the incorporation of the matrix factorization during the training process   to boost the performance of DeepMP.

\section{ACKNOWLEDGEMENT}

This work was supported by the Engineering and Physical Sciences Research Council (EPSRC) Grant numbers EP/S000631/1 and EP/K014277/1 and the MOD University Defence Research Collaboration (UDRC) in Signal Processing.

\end{document}